\documentclass[12pt]{article}
\makeatletter
\def\section{\@startsection {section}{1}{\z@}{-3.5ex plus -1ex minus
 -.2ex}{2.3ex plus .2ex}{\large\bf}}
\def\subsection{\@startsection{subsection}{2}{\z@}{-3.25ex plus -1ex 
minus -.2ex}{1.5ex plus .2ex}{\normalsize\bf}}

\newcommand{\be}{\begin{equation}}
\newcommand{\ee}{\end{equation}}

\def\inbar{\vrule height1.5ex width.4pt depth0pt}
\def\IC{\relax\,\hbox{$\inbar\kern-.3em{\rm C}$}}
\def\IG{\relax\,\hbox{$\inbar\kern-.3em{\rm G}$}} \def\IB{\relax{\rm
I\kern-.18em B}} \def\ID{\relax{\rm I\kern-.18em D}}
\def\IL{\relax{\rm I\kern-.18em L}} \def\IF{\relax{\rm I\kern-.18em
F}} \def\IH{\relax{\rm I\kern-.18em H}} \def\II{\relax{\rm
I\kern-.17em I}} \def\IN{\relax{\rm I\kern-.18em N}}
\def\IP{\relax{\rm I\kern-.18em P}}
\def\IQ{\relax\,\hbox{$\inbar\kern-.3em{\rm Q}$}}
\def\bfzero{\relax\,\hbox{$\inbar\kern-.3em{\rm 0}$}}
\def\IK{\relax{\rm I\kern-.18em K}}
\def\IG{\relax\,\hbox{$\inbar\kern-.3em{\rm G}$}} \font\cmss=cmss10
\font\cmsss=cmss10 at 7pt 
\def\IR{\relax{\rm I\kern-.18em R}}
\def\ZZ{\relax\ifmmode\mathchoice {\hbox{\cmss Z\kern-.4em
Z}}{\hbox{\cmss Z\kern-.4em Z}} {\lower.9pt\hbox{\cmsss Z\kern-.4em
Z}} {\lower1.2pt\hbox{\cmsss Z\kern-.4em Z}}\else{\cmss Z\kern-.4em
Z}\fi} \def\bfone{\relax{\rm 1\kern-.35em 1}}   
  
 \def\bar{\overline} 
\begin{document}
%\begin{titlepage}
\rightline{NORDITA-2000/125 HE} 
\rightline{SPIN-2000/36}
\vskip 1.8cm
\centerline{\Large \bf How much can supergravity teach us} 
\vskip 0.4cm
\centerline{\Large \bf  about the microscopic features of BPS
black holes?}
\vskip 1.2cm 
\centerline{\bf M. Bertolini $^a$ and M. Trigiante $^{b}$}
\vskip .8cm \centerline{\sl $^a$ NORDITA,
Blegdamsvej 17, DK-2100 Copenhagen, Denmark}
\vskip .4cm \centerline{\sl $^{b}$ Spinoza Institute, Minnaert Building, 
Leuvenlaan 4, 3584 CE Utrecht, The Netherlands}
\vskip 1cm
\begin{abstract}
We review recent results in the study of regular four dimensional BPS black 
holes in toroidally compactified type II (or M) theory. We discuss the generating 
solution for this kind of black holes, its microscopic description(s), and compute 
the corresponding microscopic entropy. These achievements, which 
provide a description 
of the fundamental degrees of freedom accounting for the entropy of any regular
 BPS black hole in the theory under consideration, are inscribed 
within a research project aimed to the study of the microscopic properties 
of this kind of solutions in relation to U--duality invariants computed on the corresponding macroscopic 
(supergravity) description. 
\end{abstract}
%\end{titlepage}
%\renewcommand{\thefootnote}{\arabic{footnote}}
%\setcounter{footnote}{0} \setcounter{page}{1} 

%%%%%%%%%%%%%%%%%%%%%%%%%%%%%%%%%%%%%%%%%%%%%%%%%%%%%%%%%%%%%%%%%%%%%%
%\tableofcontents  
\vskip 0.7cm
\section{Introduction}
One of the main issues of the ``second string revolution''
(1995) is the concept of string dualities which provided a new insight
into the  non--perturbative side of the known superstring
theories. These  dualities are mappings between regimes of different
superstring  theories (some of them have been  verified while other
just conjectured). Their  existence naturally induces  to consider the
known superstring theories as  perturbative realizations on different
backgrounds of a fundamental  theory of gravity (FTG) whose general
formulation however is still missing.  It is known that the low energy
limit of superstring theory is described by supergravity. Although
supergravity in this picture is regarded just as a  {\it macroscopic}
theory, it is expected to possess important informations about the
FTG. Indeed, it has been argued \cite{towhull}  that the largest
(continuous) global symmetry  group $U$ of the supergravity field
equations and Bianchi identities at classical level should encode the
definition, as a suitable discrete group $U(Z)$,  of  the conjectured
superstring $U$--duality, namely the ultimate  duality connecting all
superstring theories realized on various backgrounds.  This duality is
thus expected to be an exact symmetry of the FTG.  Unfortunately not
much is known about the group $U(Z)$, starting from the  very
definition and its action on superstring states. On the other hand the
action of the group $U$ on the supergravity solutions is, in
principle, known.

\begin{figure}[t]
\label{Uu}
\vskip 20pt
\input epsf
\epsfxsize=250pt
\centerline{\epsffile{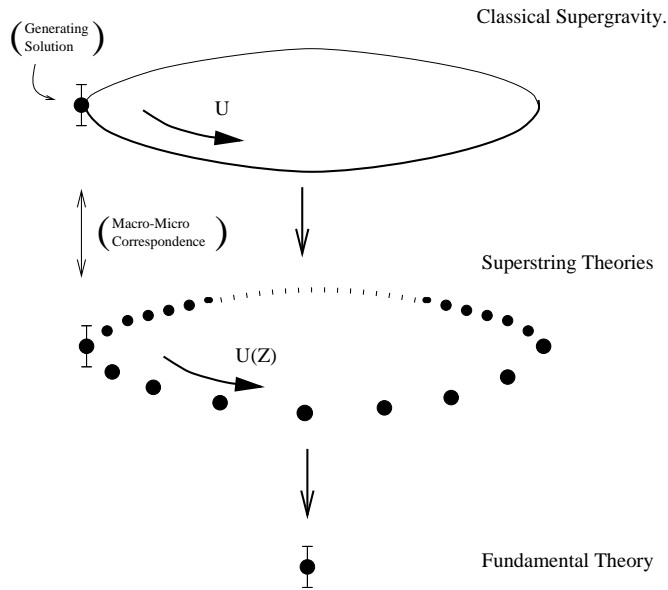}}
\caption{The action of classical and quantum $U$--duality.}
\end{figure}

A fundamental role in probing superstring dualities has been played so
far by the BPS black hole solutions of supergravity. These solutions
are characterized by the property of preserving a fraction of the
original supersymmetries, and this feature protects their physical
quantities, to a certain extent, from quantum corrections. As a
consequence of their supersymmetry, BPS black holes in supergravity
are expected to correspond to exact solutions of superstring
theory. The BPS condition moreover is $U$--duality invariant. This
allows to characterize these supergravity solutions within orbits of
the continuous  $U$--duality group, defined by a certain number of
$U$--invariants  $\{{\cal I}_k\}$ (e.g. the entropy). All the physical
properties of the BPS solutions entering the  same $U$--duality orbit
are expected to be encoded in the corresponding {\it generating
solution}. The generating solution of BPS black holes is defined,
within a certain supergravity theory, as the solution depending on the
least number of parameters such that the invariants $\{{\cal I}_k\}$
are free for a certain choice of the boundary conditions. As a
consequence of its definition, by acting on the generating solution by
means of $U$ one recovers the whole $U$--duality orbit. A suitable
discrete set of points within this orbit should correspond to
superstring black holes (non--perturbative solutions) connected by the
action of $U(Z)$ and which therefore represent different descriptions
of a same solution within the FTG (see figure \ref{Uu}). 
The microscopic degrees of freedom described by the FTG are indeed
related to  invariants of the group $U(Z)$. Pinpointing the {\it
exact} correspondence between the macroscopic (supergravity) and
microscopic descriptions  (e.g. in terms of D--branes in a suitable
regime) of a generating solution, one would in principle be able to
study systematically the microscopic realization of a generic solution
in the same orbit. Moreover this could be the  first step in order to
unravel the action of  $U(Z)$ on stringy objects in higher dimensions
and to ultimately deduce their fundamental degrees of freedom.

Here we review some recent results achieved in \cite{mm1,mm2,micro} 
where a macroscopic (supergravity) starting point was adopted for a 
systematic microscopic analysis of regular BPS (static, spherically 
symmetric) black holes within type II (M) theory compactified down to 
four dimensions on tori, and whose zero modes are described by $N=8$ 
four dimensional supergravity.

The paper is organized as follows. In section 2 we shall start addressing 
the question: {\it how much can we learn at classical supergravity
level about the microscopic description of a BPS solution?} A possible
answer will lead us to discuss the mathematical analysis carried out in
\cite{mm1} which provides an intrinsic group theoretical
characterization of the scalar and vector fields  in the $D=4,N=8$
theory in terms of dimensionally reduced type II fields. The
geometrical framework so defined turns out to provide the convenient
``laboratory'' in which to systematically study the microscopic
descriptions of BPS solutions and their duality relations.  Using
these tools one can then characterize R--R charged generating
solutions of regular BPS black holes as elements of a suitable
equivalence class defined with respect to the action of $S$ and $T$
dualities. This result is discussed in section 3 and allows us to
formulate the precise correspondence, worked out in \cite{mm2},
between the parameters defining R--R charged (D--brane) 
microscopic descriptions
of black hole solutions and the supergravity quantities  related to
its macroscopic descriptions. This will be used, in section  4, for
providing a type IIA/IIB/M--theory description of the generating
solution of regular BPS black holes, and a prediction on the
expression of  the macroscopic entropy (at tree level) in terms of
microscopic parameters. Finally, in section 5, focusing on the
M-theory description of the generating solution, the same expression for
the entropy will be retrieved from a counting of BPS micro--states, 
extending  the analysis in \cite{msw} to the  toroidal case. This last result 
was achieved in \cite{micro}. From the very definition of generating
solution, this analysis accounts for the microscopic entropy of the
most general black hole solution of this kind.

%%%%%%%%%%%%%%%%%%%%%%%%%%%%%%%%%%%%%%%%%%%%%%%%%%%%%%%%%%%%%%%%%%%%%
\section{Supergravity Laboratory}
The only prediction which may be drawn at classical supergravity level
on the microscopic description of a BPS solution is clearly limited
to the background fields which couple to it.  This can be done
for instance by associating each superstring scalar and vector
zero--mode with quantities intrinsic to the $U$--duality group of the
low--energy supergravity \cite{solv2,mm1}.

The $D=4,N=8$ supergravity is a maximally extended supersymmetric
theory, i.e. it has $32$ supercharges. Its bosonic sector consists of
the graviton, $70$ scalar fields, spanning the homogeneous manifold ${\cal
M}_{scal}=E_{7(7)}/SU(8)$, and 28 vector fields. The latter are
related to a vector of $56$ {\it quantized charges} $(p^\Lambda,
q_\Sigma)$, which transforms in the $Sp(56)$ of $E_{7(7)}$, and a {\it
central charge} matrix $Z_{AB}$ entering the local realization on the
moduli space of the supersymmetry algebra and transforming in the
${\bf 28}$ of $SU(8)$. The former charges are moduli--independent and
should be regarded just  as supergravity parameters, while the latter
are moduli dependent  and are related to the physical charges,
i.e. the actual charges one would measure in the asymptotically flat
radial infinity of a black hole solution.

The $U$--duality group of the classical theory is $U=E_{7(7)}$
\cite{cremjul}.  It acts as a generalized electro--magnetic duality,
i.e. it has a non--linear action on the scalar fields and a linear
(symplectic)  action on the vector of quantized charges. As previously
mentioned, the  $D=4,N=8$ theory describes the low--energy limit of
type II superstring  theory on $T_6$  (or M--theory on $T_7$). The
first step towards a group  theoretical characterization of the
ten--dimensional origin of the scalars  and charges in this
supergravity model is to use a linear algebraic description of the
scalar fields. This is achieved by adopting the {\it solvable Lie
algebra} (SLA) parameterization of the scalar manifold
\cite{solv2,n2,solv1}, which consists in describing the scalar fields
as local parameters of a solvable Lie algebra which generates
(globally) the scalar manifold as a solvable Lie group. Homogeneous
non--compact manifolds of symmetric type like ${\cal M}_{scal}$ do
admit such a representation:
\begin{equation}
{\cal M}_{scal}={\rm Exp}(Solv(U))
\end{equation}
The algebra $Solv(U)$ is defined by the Iwasawa decomposition of
$E_{7(7)}$ and  can be written as $Solv(U)={\cal C}\oplus {\cal N}$,
where ${\cal C}$ is the Cartan subalgebra of $E_{7(7)}$ while ${\cal
N}$ is the nilpotent subalgebra of $E_{7(7)}$ generated by all the shift
generators corresponding to positive roots.  In this framework a one
to one correspondence between the scalar fields and the generators of
$Solv$ is defined.

Two relevant duality groups for our discussion are the $S=SL(2,R)$ and
$T=O(6,6)$ subgroups of $U$, defined as the continuous counterparts at
the  classical level of the discrete $S$ and $T$ superstring
dualities\footnote{In our formalism, this $S$--duality has not to be confused 
with the self--duality of the type IIB theory. Indeed in four dimensions it acts only 
on the effective dilaton and the four dimensional axion deriving from the 
NS-NS Kalb--Ramond field. The present $SL(2,R)$ has just a $O(1,1)$ intersection with the 
ten dimenional type IIB $SL(2,R)$ symmetry group.}. Since these dualities are the 
largest preserving the R--R and NS--NS identities of  the fields, decomposing 
$Solv(U)$ with respect to $Solv(S)\times Solv(T)$ one  may achieve an intrinsic
characterization of the R--R and NS--NS fields at  classical
supergravity level. On the other hand  the dimensional reduction of
type II  superstring to four dimensions may be performed through
intermediate steps which define, in the low--energy limit, higher
dimensional maximal  supergravities, with their own $U$--duality group
$U_{D>4}$ at tree level.  Fixing then the embedding of $Solv(U_{D>4})$ within
$Solv(U)$ for various $D>4$ allows to identify in a consistent way the
scalar fields of the $N=8$ theory, as associated with the
corresponding generators of $Solv(U)$, with dimensionally reduced type
II zero--modes. 

On the vector field side, it is convenient to work with a set of
 physical charges $(y^\Lambda, x_\Sigma)$ (transforming under $SU(8)$)
 which are expressed in the same basis of weights $\{\vec{\lambda}\}$,
 generating the ${\bf 56}$ of $U$, as the quantized charges
 $(p,q)$. These charges are obtained from the vector  $({\rm
 Re}Z_{AB},{\rm Im}Z_{AB} )$ through a suitable  rotation and are
 related to the quantized charged $(p,q)$ by a  {\it
 moduli--dependent} symplectic transformation which makes them
 quantized as well \cite{mm1}.   Decomposing the weight basis
 $\{\vec{\lambda}\}$ with respect to the action of the higher
 dimensional $U$--dualities $U_{D>4}$ it was possible to associate
 consistently with each weight $\vec{\lambda}$ a one--form electric or
 magnetic potential in  four dimensions deriving from suitable ten
 dimensional type II zero--modes.

As a result of this first group
 theoretical analysis an $N=8$ {\it algebraic dictionary} \cite{mm1}
 could be established on the weight lattice $\Lambda_{W}(U)$ of $U$ in
 which the {\it directions} (namely Cartan generators in ${\cal C}$)
 and the {\it positive roots} are associated with scalar fields
 (through the SLA parameterization) and the weights
 $\{\vec{\lambda}\}$ with electric and magnetic one--form potentials,
 each of these fields having a specific ten dimensional
 characterization.

%%%%%%%%%%%%%%%%%%%%%%%%%%%%%%%%%%%%%%%%%%%%%%%%%%%%%%%%%%%%%%%%%%%%%%%%%
\section{Regular BPS black holes with R--R charge}

Regular BPS black holes are BPS solutions having a finite horizon
area. They where shown to preserve  $1/8$ of the original $N=8$
supersymmetries and to interpolate between an $N=2$ vacuum of the form
$AdS_2\times S_2$ near the horizon ($r\rightarrow 0$) and a Minkowski
vacuum at radial infinity ($r\rightarrow \infty $) \cite{ferkal}. 
The physical charges of a BPS black hole solution, as previously
mentioned, are related to the (antisymmetric) central charge matrix
$Z_{AB}$ which depends on the point on the moduli space $\phi_0$,
representing the boundary condition at radial infinity of the scalar
fields,  as well as on the quantized charges.  The $U$--duality
invariants $\{{\cal I}_k\}$ of the solution are given by all the
$SU(8)$ invariants which can be built out of $Z_{AB}$. Indeed, acting
by means of a $U$--duality transformation on the scalar fields and the
quantized charges, the central charge matrix will transform under a
corresponding $SU(8)$ transformation. These invariants are {\it five}
and on the orbit of regular BPS black holes they are independent parameters. A
way of expressing them is in terms of the norm of the central charge
{\it skew--eigenvalues} $Z_\alpha$ ($\alpha=0,\dots,3$) and their
overall phase, i.e.  $\{{\cal I}_k\}=\{|Z_\alpha|,\,\Theta\}$.  By
suitably combining them it is possible to obtain a moduli--independent
invariant, namely the quartic invariant $J(x,y)$ of the ${\bf 56}$
of $E_{7(7)}$ (the orbits of BPS black holes have $J(x,y)\ge 0$ \cite{ferma}).
 This is the only invariant characterizing the near--horizon geometry of the 
solution. The area of the horizon  is ${\cal A}=4 \pi\sqrt{J(x,y)}$
and, using Bekenstein--Hawking formula, the tree level entropy turns
out to be \cite{ferla}: 
\begin{equation}
S={\cal A}/4=\pi\sqrt{J(x,y)}
\end{equation}
{\it The generating solution is defined by a choice of the bosonic vacuum at infinity $\phi_0$ 
and by the minimum number (i.e. five) of charges in terms of which the
invariants $\{{\cal I}_k\}$, computed on $\phi_0$, are independent
functions.} This solution can be described within a $STU$ model, which
is characterized as the smallest consistent truncation of the
$N=8$ theory on which the four $Z_\alpha$ are independent \cite{e77}. 
 The $STU$ model is an $N=2$ supergravity coupled to three vector multiplets, 
its  classical $U$--duality group $U_{STU}=SL(2,R)^3\subset U$ is defined by 
the isometry group of the scalar manifold ${\cal M}_{STU}=U_{STU}/SO(2)^3$. The latter, 
in the SLA formalism, may be described as a solvable Lie group generated by a solvable 
Lie algebra $Solv_{STU}$ which is parametrized by just three dilaton fields $b_i$ and three 
axions $a_i$. This model moreover has four vector fields which give rise to  eight 
quantized charges $(p^\alpha,\, q_\beta)$ and eight
physical charges $(y^\alpha, \,x_\beta)$.
In the light of the previously defined
 algebraic dictionary, different  microscopic descriptions of the
 generating solution can be put in  correspondence with different
 embeddings of the $STU$ model within the $N=8$ one  (defined by the
 embedding of the corresponding solvable Lie algebras and  charge
 weights\footnote{In other words, by the embedding of the weight
 lattices: $\Lambda_W(U_{STU})\subset \Lambda_W(U)$.}). 

Dualities relating different embeddings of the $STU$ model are naturally
 described in terms of the  action on $\Lambda_W(U_{STU})$ of {\it
 automorphisms} ($Aut$)  of the relevant duality algebra \cite{mm1}.
 In order to characterize the generating  solution as charged with
 respect to R--R or NS--NS fields, we would need then to consider the
 action of the $S\times T$ dualities through their authomorphism group
 ($Aut(S\times T)$). The Dynkin diagram of the $T$ algebra is
 $D_6$. It has  {\it inner} and {\it outer}  automorphisms, the latter
 being related,  through Weyl transformations, to the only symmetry of
 $D_6$ (for a study of  Weyl duality  transformations in supergravity
 see \cite{stelle}). These  outer automorphisms are particularly
 interesting since they are not a symmetry  and can be thought of as
 relating two different descriptions of the same theory, namely the
 type IIA and type IIB ones. Indeed, using the SLA representation it
 was shown in \cite{mm1} that the outer automorphisms of $T$
 correspond to a ``large $\leftrightarrow $ small radius''
 $T$--dualities along an odd number of directions inside $T_6$.
Taking into account the action of these outer automorphisms the 
$N=8$ algebraic dictionary was consistently 
enlarged to accommodate both type IIA and IIB descriptions of the $N=8$ theory
(see tables 2 and 3 of \cite{mm1}). 

Within the mathematical framework defined above, 
two $T$--dual embeddings $STU_1,\,STU_2$ of the $STU$
model, for which the charges were  related to suitable  R--R
one--forms, were worked out in \cite{mm1}. The corresponding
 two descriptions of
the fields in the $STU$ model in terms of $E_{7(7)}$ weights are
mapped into each other through an outer automorphism of $T$, which is
interpreted, in the SLA formalism, as a ``large $\leftrightarrow$
small radius'' duality along the directions  $x^5,x^7,x^9$ of
$T_6$ (in our notation the compact directions are $x^4,\dots ,\,x^9$
while the non--compact are $x^0,\dots,\,x^3$). One embedding ($STU_1$)
can be indeed consistently described in the type IIA setting while the
other  ($STU_2$) in the type IIB one. In particular, from the $N=8$
algebraic dictionary, it is possible to characterize the {\it axions} 
of the $STU_1$ embedding as deriving from the antisymmetric tensor 
$B_{MN}$ ($\{a_i\}=\{B_{45},\,B_{67},\,B_{89}\}$) while those in 
$STU_2$ as deriving from the metric $G_{MN}$
($\{a_i\}=\{G_{45},\,G_{67},\,G_{89}\}$). As far as the vector fields
are concerned, in an analogous way the charges $(y^\alpha,x_\beta)$ in
the type IIA embedding $STU_1$ are associated with 1--form (magnetic
and electric) potentials deriving from the following components of the
ten dimensional R-R fields $A_M,\,A_{MNP}$: {\small
\begin{eqnarray} (y^\alpha)&\leftrightarrow& (A_{\mu 456789},A_{\mu
6789},A_{\mu 4589},A_{\mu 4567})\nonumber\\  (x_\beta)
&\leftrightarrow& (A_{\mu},A_{\mu 45},A_{\mu 67},A_{\mu 89})
\label{iia}
\end{eqnarray}}
while for the type IIB embedding $STU_2$ this correspondence between
charges  and components of the R--R forms $A_{MN},\, A_{MNPQ}$ reads:
{\small \begin{eqnarray} (y^\alpha)&\leftrightarrow& (A_{\mu
468},A_{\mu 568},A_{\mu 478},A_{\mu 469})\nonumber\\  (x_\beta)
&\leftrightarrow& (A_{\mu 579},A_{\mu 479},A_{\mu 569},A_{\mu 578})
\label{iib}
\end{eqnarray}} 
From this background field prediction and from the values of the
physical charges of the generating solution  at infinity (for a
suitable choice of the boundary conditions), two  $T$--dual D--brane
descriptions, corresponding to the embeddings discussed above, can be
consistently worked out and precise relations  established between the
parameters defining the macroscopic (supergravity) and microscopic
(D--brane) descriptions of the generating solution
\cite{mm2}.  Finally, acting on $STU_{1,2}$ by means of $Aut(S\times
T)$ one could define an equivalence class of R--R charged embeddings
of the $STU$  model (yielding all the R--R charged generating
solutions) within the $N=8$ theory.

%%%%%%%%%%%%%%%%%%%%%%%%%%%%%%%%%%%%%%%%%%%%%%%%%%%%%%%%%%%%%%%%%%%%%%%%%%%%%%%%
\section{Generating solution in type IIA/IIB/M--theory}
The machinery reviewed in the previous section was used in \cite{mm2} to construct 
out of the macroscopic description of the  generating solution, possible IIA, IIB and 
M-theory microscopic realizations.  Let us briefly summarize its structure.
The ansatze for the generating solution in terms of metric, scalar fields 
$z^i=a_i+{\rm i}\,b_i$ and the four vector field strengths is:
\begin{eqnarray}
ds^2&=&e^{2{\cal U}\left(r\right)}dt^2-e^{-2{\cal
U}\left(r\right)}d\vec{x}^2 ~~~~~
\left(r^2=\vec{x}^2\right)\nonumber\\  z^i(x)&=&z^i(r) \nonumber \\
F^\Lambda(r)&=&\frac{p_\Lambda}{2 r^3}\epsilon_{krs} x^k dx^r \wedge
x^s - \frac{l_\Lambda(r)}{r^3} e^{2{\cal U}(r)} dt \wedge \vec{x}
\cdot d\vec{x} \quad,\quad \Lambda=0,1,2,3
\label{fishlicense}
\end{eqnarray}
$l_\Lambda$ being the moduli--dependent electric charges defined in
\cite{bft}.  The solution of both field equations and the first order
differential equations representing the BPS condition can be expressed
in terms of harmonic functions.  In particular the generating solution
we are interested in will depend only on five charges chosen in such a
way that the three charges which are set  to zero break completely the
$SO(2)^3$ local symmetry of the model.  A possible choice for these
charges is $p^1,\,p^2,\,p^3,\,q_0,\,q_1$.  Secondly we choose the
following boundary condition for the scalar fields at radial infinity:
\begin{eqnarray}
(\phi_0) &\equiv &\cases{a_1=a_2=0;\,a_3=g\cr b_i=-1}\nonumber\\
g&=&\frac{q_1}{p^1+p^2}
\label{phinfty}
\end{eqnarray}
This allows to write the physical charges $(x,y)$ in terms of the
moduli--independent ones $(p,q)$  and to use the former to describe
the solution. The symplectic transformation connecting the two sets
of charges on our solution is: $y^0=0,\,
y^i=p^i,\,x_0=q_0,\,x_1=-x_2=g\,p^1$.  Let us now introduce the
following five harmonic functions:
\begin{eqnarray}
&&H^i(r)\,=\,1+\frac{\sqrt{2}y^i}{r} \quad \mbox{with}\quad i=1,2,3
\nonumber \\ &&H_0(r)\,=\,1+\frac{\sqrt{2}x_0}{r} \quad
\mbox{and}\quad\nonumber \\
&&H_1(r)\,=\,g\,\left(H^1(r)+H^2(r)-1\right)
\label{armf}
\end{eqnarray}
where the constant $g$, which is fixed by supersymmetry, is given in
eq.(\ref{phinfty}) and  can be alternatively expressed in terms of
the physical charges as $g=x_1/y^1$. The solution as far as the scalar
fields and metric are concerned has the following form:
\begin{eqnarray}
&&a_1\,=\,\frac{-H_1H^1+gH^2}{2 H^2H^3}\;,\;\; b_1\,=\,
-\sqrt{\frac{H_0H^1}{H^2H^3}-\frac14\left(\frac{H_1H^1-gH^2}{H^2H^3}\right)^2}\nonumber
\\ &&a_2\,=\,\frac{H_1H^1-gH^2}{2 H^1H^3}\;,\;\; b_2\,=\,
-\sqrt{\frac{H_0H^2}{H^1H^3}-\frac14\left(\frac{H_1H^1-gH^2}{H^1H^3}\right)^2}\nonumber
\\ &&a_3\,=\,\frac{H_1H^1+gH^2}{2 H^1H^2}\;,\;\; b_3\,=\,
-\sqrt{\frac{H_0H^3}{H^1H^2}-\frac14\left(\frac{H_1H^1-gH^2}{H^1H^2}\right)^2}\nonumber
\\ &&{\cal U}\,=\,-\frac{1}{4}\ln\left(H_0H^1H^2H^3
-\frac{1}{4}(H_1H^1-gH^2)^2\right)
\label{ans}
\end{eqnarray}
We can see from the above expressions that the parameter $x=x_1=-x_2$
has a special role: switching it off the axion fields $a_i$ become
identically  zero and the solution reduces to a four parameter purely
dilatonic one. We shall comment in  the sequel on the microscopic
interpretation of this {\it fifth} parameter.

Starting from the two IIA and IIB $T$--dual embeddings previously
defined, it was possible to characterize two microscopic descriptions
of the generating solution where all the  four one--form potentials 
derived from R--R ten dimensional fields. In this way the solution 
could be described in the weak string coupling limit in terms of bound 
states of D--branes wrapped on $T_6$. 
On the type IIB front the microscopic system consists of $N_0$, $N_1$, $N_2$, 
$N_3$ D3--branes arranged within $T_6$ in such a way as to preserve $N=1$ 
supersymmetry and this requires the relative rotation between each couple of
D3--brane to be a $SU(3)$ rotation. The corresponding $T$--dual type
IIA system consists of a set of D0--branes and three sets of
D4--branes along the four--cycles $(6789)$, $(4589)$ and $(4567)$. In
addition, there is a  magnetic flux switched on the world volume of
the latter (i.e. along $(4567)$) which is proportional to a rational
number $\gamma=m/n$, where the integers $m,n$ are related to the
non-trivial angle $\theta$ characterizing  the type IIB configuration
by the condition: $n\, {\rm sin}\,\theta=m\,  {\rm cos}\,\theta$. This
flux induces effective D0 and D2 charges via Chern-Simons couplings \cite{bec}. 
The  eleven dimensional $S$--dual (M--theory) correspondent of the type 
IIA configuration is summarized in table 1 and consists
of a set of three bunches of M5--branes intersecting on a  (compact)
line  and with  non--trivial 3--form field strength $h^{(3)}=db^{(2)}$
switched--on on their world volume. In this phase the compact space has 
an extra dimension, of course, $T_6\times S_1$. Besides the M5--branes, 
which are $N_1,N_2,N_3\,n^2$ respectively, there are $N_0+N_3\,m^2$ units of KK
momentum  along the spatial $10^{th}$ direction and the magnetic flux,
related to non--trivial 3--form field strength excited on the
M5--brane, is proportional to $\gamma$. This is of course the same flux 
present in the $S$--dual type IIA description.  
{\small 
\begin{table} [ht]
\begin{center}
\begin{tabular}{|c|c|c|c|c|c|c|c|c|c|c|c|}
\hline  Brane & 0 & 1 & 2 & 3 & 4 & 5 & 6 & 7 & 8 & 9 & 10 \\ \hline
$P_L$     & $\cdot$  &$\cdot$ &$\cdot$ &$\cdot$ &$\cdot$ &$\cdot$
&$\cdot$ &$\cdot$ &$\cdot$ &$\cdot$ & $\times$ \\   \hline $M5$ &
$\times$ &$\cdot$ &$\cdot$ &$\cdot$ &$\cdot$ &$\cdot$ &$\times$
&$\times$ &$\times$  &$\times$ & $\times$  \\   \hline  $M5$ &
$\times$ &$\cdot$ &$\cdot$ &$\cdot$ &$\times$ &$\times$ &$\cdot$
&$\cdot$ &$\times$  &$\times$ & $\times$  \\   \hline $M5+h^{(3)}$ &
$\times$ &$\cdot$ &$\cdot$ &$\cdot$ &$\times$ &$\times$ &$\times$
&$\times$ &$\cdot$   &$\cdot$ & $\times$  \\ \hline
\end{tabular}
\caption{The M--theory generating configuration.}
\end{center}
\label{Mbrane}
\end{table}}

The precise relation between the macroscopic charges $(y,x)$ as related 
to the effective charges along the various
cycles  of $T_6\times S_1$ and the microscopic parameters
$\{N_\alpha,p,q\}$, is given in table \ref{Mbrane2}.
{\small 
\begin{table} [ht]
\begin{center}
\begin{tabular}{|c|c|c|c|c|}
\hline  M-brane cycles & Type IIA cycles & Type IIB cycles & Charges & 4D Charges  \\
\hline $KK$--monopole & $D6$(456789) & D3(468) & $0$ & $y_0$\\    \hline
$M5$(6789$|$10)& $D4$(6789) & $D3$(568) & $N_1$ & $y_1$ \\    \hline
$M5$(4589$|$10)& $D4$(4589) & $D3$(478) & $N_2$ & $y_2$ \\  \hline
$M5$(4567$|$10)& $D4$(4567) & $D3$(469) & $N_3 n^2$ & $y_3$ \\    \hline
$KK$--momentum &$D0$ & D3(579) & $N_0 + m^2 \; N_3$ & $x_0$ \\    \hline
effective $M2$(45)&effective $D2$(45) & $D3$(479) & $- m\, n \;N_3$ & $x_1$ \\
\hline effective $M2$(67)&effective $D2$(67) & $D3$(569) & $m\, n\;N_3$ & $x_2$ \\
\hline effective $M2$(89)&effective $D2$(89) & $D3$(578) & $0$ & $x_3$ \\    \hline
\end{tabular}
\end{center}
\caption{The relation between M/IIA/IIB microscopic parameters and the macroscopic charges.}
\label{Mbrane2}
\end{table}}

One of the crucial issues in the stringy description of supergravity
black holes is the  precise characterization of the parameters
entering the solution in terms of microscopic
quantities. In particular, the interpretation of the fifth parameter
of  the generating solution, which is related to a non trivial overall 
phase $\Theta$ of the central charge skew--eigenvalues, is rather tricky
to be dealt with \cite{hull,bala}. Thanks to the construction reviewed
in the previous section we can clearly understand its role. It is clear from 
eq.(8) that the non vanishing of this parameter is related to our 
solution being an axionic one. Switching it
off indeed, as previously noticed, 
one gets back a pure (four parameter) dilatonic solution. 
The number of  independent
harmonic functions is four, in both cases. This is an expected feature 
and is related to the fact  that within the five invariants  of the
$U$--duality group four are moduli--dependent and one is
moduli--independent (namely the $J(x,y)$ polynomial related to the entropy). 
The conclusion which can
 be drawn is thus that the generating solution is intrinsically axionic, i.e.,
in order to recover the full $U$--duality orbit the pure dilatonic solution is not
sufficient. According to table \ref{Mbrane2} one can also easily understand
the microscopic interpretation of this parameter: it is related to the
non-trivial magnetic flux switched-on on the D4 or M5-brane world-volumes, 
in the type IIA and M--theory phases, respectively. From a type IIB view
point, the same quantity is related to a non--trivial $SU(3)$ 
rotation between couples of intersecting D3-branes.
 Switching it off one gets back a
 configuration of orthogonally intersecting D3--branes or a type IIA/M brane
configuration without any non-trivial world--volume field. This is consistent 
with the interpretation of the axion fields in the corresponding 
 two embeddings  $STU_2$ and $STU_1$ discussed in
section 3: in the former they are related to non--diagonal  components of
the ten dimensional metric tensor, in the latter to internal components
of  the Kalb--Ramond field. \par
Using Bekenstein--Hawking formula on the generating solution, eq.s 
(\ref{fishlicense})--(\ref{ans}), 
and expressing the charges at infinity in terms of  the microscopic
parameters by means of table \ref{Mbrane2}, the following expression
for the macroscopic entropy (tree level) in terms of the microscopic
quantities can be derived:
\begin{equation}
S = 2\,\pi\,\sqrt{N_1N_2N_3\, n^2\left[N_0 + m^2\, N_3- \frac{1}{4}
m^2\, N_3\frac{\left(N_1+N_2\right)^2}{N_1N_2}\right]} 
\label{Smicro}
\end{equation}
In the following, we shall review the derivation of the expression
(\ref{Smicro}) from a microscopic BPS state counting.

%%%%%%%%%%%%%%%%%%%%%%%%%%%%%%%%%%%%%%%%%%%%%%%%%%%%%%%%%%%%%%%%%
\section{Microscopic entropy counting}
In \cite{micro} the M--theory description of the generating solution
was considered and its microscopic entropy computed.
Since the physical quantities related to the BPS microstates that  we
are interested in  are insensitive to smooth deformations of the
background moduli, the latter can be chosen in such a way as to make 
the microscopic entropy counting feasible. The choice made in \cite{micro}
corresponds to the regime in which the dynamics of the branes decouple from 
the bulk and their ``thickness'' is much smaller than all the other length
scales in the theory (in M--theory, the Planck length and the size of
the internal manifold $T_7$) and therefore the supergravity
description  of the solution cannot in general be trusted. In this
limit the low energy effective theory on the world volume of each
M5--brane in table \ref{Mbrane}  in the background of the other two
is a $(0,2)$ SCFT.  The quantization of the eleven dimensional system 
is therefore performed in the framework of
M--theory on $\IR^{1,3}\times T_7$ extending to the toroidal case the
results of \cite{msw} for the case of M--theory on $\IR^{1,3}\times
CY_3\times S_1$. 

A fruitful strategy for performing a microscopic
entropy counting on a BPS solution has been so far to restrict to a
particular background on which the low energy dynamics of the system
is actually described by a $1+1$ SCFT \cite{sv,v1,msw}\footnote{For a review
on microscopic entropy counting see \cite{malda,amanda,mooa} and references therein.}. 
In this limit, the
asymptotic value of the degeneracy of states for a  high excitation
level is given by the Cardy formula \cite{carlip}, which, if we
restrict only to the left--movers (see below), has the form:
\begin{eqnarray}
\rho(n)\approx e^{2\pi\sqrt{c_L\,h/6}}
\label{cardy}
\end{eqnarray}
$\rho(h)$ being the state degeneracy for the left--moving excitation
level $h$ while $c_L$ is the central charge of the left--mover sector
(the above formula holds in the limit $h\gg c_L$). Using Boltzman
equation and eq. (\ref{cardy}) the microscopic entropy can be
expressed as:
\begin{eqnarray}
S=\ln \rho(h)\approx 2\pi\sqrt{\frac{c_L\,h}{6}}
\label{entropy}
\end{eqnarray}
As far as the M--brane
system in table \ref{Mbrane} is concerned the effective
low energy description on terms of a $1+1$ SCFT is obtained in the
limit in which the radius R of the eleventh dimension $S_1$ is
much larger than the linear size of the orthogonal $T_6$. In
particular, the M5 branes can be described as wrapped on $P\times
S_1$, $P$ being an holomorphic cycle of $T_6$ (seen as a complex
3--manifold). As the size of $P$ shrinks with respect to R the low
energy dynamics of the M5 brane is described by the dimensional
reduction of the $(0,2)$ SCFT to $S_1\times \IR$, which is a $(0,4)$
SCFT in $1+1$ dimensions.  The bosonic fields of the latter theory are
the moduli of the cycles $P$ and of the chiral two form $b^{(2)}$. BPS
states in this framework are annihilated by all the four right--moving
supercharges and therefore are characterized only  by the left--moving
excitation level $h$: $|BPS>=\vert h\rangle_L \otimes \vert
\bar{0}\rangle_R$. 

In order to use equation (\ref{entropy}) for
computing the entropy, we need therefore to determine $c_L$ and $h$ in
terms of the charges $(x,y)$ characterizing our black hole.
%%%%%%%%%%%%%%%%%%%%%%%%%%%%%%%%%%%%%%%%%%%%%%%%%%%%%%%%%%%%%%%%%%%%%
\subsection{Computation of $c_L$}
The general expression for $c_L$ is  $c_L=N_L^B+N_L^F/2$, where
$N_L^B$ and $N_L^F$ are the number of bosonic and fermionic degrees of
freedom in the left--moving sector, respectively. The former consists
of the left--moving moduli $d_P$  of $P$ and of the moduli associated to the 
form $b^{(2)}$. Let us outline how to evaluate their contribution.

We may associate $P$ with its fundamental class
$\left[P\right]\in H^{(2)}(T_6,\ZZ)$. In our configuration
$\left[P\right]$ is expanded in a system of three 2--cycles $\alpha_i$
dual to each  four cycle in $P$ as in table \ref{Mbrane}
($\alpha_i\equiv dx^a\wedge dx^b$, $\{(a,b)\}\equiv
\{(4,5),(6,7),(8,9)\}$) and the corresponding coefficient is the
integer magnetic charge $y^i$: $\left[P\right]=\sum_i\,
y^i\,\alpha_i$. The intersection matrix restricted to $(\alpha_i)$ is
denoted by $D_{ijk}=(\int_{T_6}\,\alpha_i\wedge\alpha_j\wedge\alpha_k)/6$ 
and is a symmetric matrix whose only non vanishing entry is
$D_{123}=1/6$. The volume of $P$ is therefore $\mbox{Vol}(P)=\int_{T_6}\,
\left[P\right]^3=6 \,D_{ijk}y^iy^jy^k= 6y^1 y^2 y^3$. The moduli of
$P$ are, roughly speaking, the number of ways $P$ can be deformed
leaving the magnetic charges (i.e. $\left[P\right]$) fixed. Using
tools of  algebraic  geometry  \cite{griff}, the number of these
holomorphic  deformations can be exactly determined,
 assuming $P$ to be a {\it very ample divisor} \cite{msw}, and turns out to
be in our  case: $d_P=\mbox{Vol}(P)/3-2$.\footnote{ The divisor $P$  
may be indeed characterized as the zero locus of a
holomorphic section of a line bundle ${\cal L}$ on $T_6$. This section
is defined up to multiplication by a non vanishing complex number. Any
other holomorphic section of ${\cal L}$ will define through its
zero--locus a different divisor $P^\prime$ (a deformation of $P$) with
the same fundamental class: $\left[P^\prime\right]=\left[P\right]$.
Therefore the space of all the holomorphic deformations of $P$ with
this property is a linear space (complete linear system) which
coincides with the projectivization  of the space of holomorphic
sections of ${\cal L}$: $\IP\left[H^{(0)}(T_6,{\cal O}({\cal
L}))\right]\sim \IC \IP^{d_P/2}$. Assuming $P$ to be 
a very ample divisor the higher order cohomology groups 
$H^{(n)}(T_6,{\cal O}({\cal L}))$ become trivial and $d_P$ can be computed 
as an index, yielding 
the above result.}

As far as the moduli of $b^{(2)}$ are concerned, indicizing  by $a$
the coordinates on $P$ and by $\beta$ those on $S_1\times \IR$, the
dimensional reduction of $b^{(2)}$ yields the non trivial fields
$b_{ab}$ and $b_{a\beta}$  in the  $1+1$ theory. The former split into
self--dual $b_{ab}^+$ and anti--self--dual components $b_{ab}^-$ on
$P$ (spanning the spaces $b^\pm$ respectively) which, as a consequence
of the selfduality of $h^{(3)}$ are associated with the  left--moving
and right--moving moduli, respectively. The dimensions of $b^\pm$ can be 
easily 
determined using the Hodge index theorem (again under the hypothesis of $P$
very ample). The components $b_{a\beta}$, on which we shall comment in a moment, 
are non--dynamical vector fields spanning a space  of dimension 
$b_1(P)=2\, h^{(1,0)}(P)$. 

The number of the left--moving and right--moving fermionic moduli,
$N_L^F$ and $N_R^F$, can be shown to be related to the dimensions of
the cohomology groups $H^{(2r+1,0)}(P)$ and $H^{(2r,0)}(P)$. The final 
expression for the various quantities cited so far turns out to be the 
following one:
\begin{eqnarray}
\label{bf}
N_L^B&=&\int_{T_6} \left[ P\right]^3 + 2 h_{(1,0)} \,\,\,\{- 2 h_{(1,0)} =
\int_{T_6} \left[ P\right]^3\} \nonumber \\ N_L^F&=&4 h_{(1,0)} \,\,\,\{- 4 h_{(1,0)} =
0\} \nonumber \\  N_R^B&=&\frac{2}{3}\int_{T_6} \left[ P\right]^3 + 2 h_{(1,0)}
\,\,\,\{- 2 h_{(1,0)} = \frac{2}{3}\int_{T_6} \left[ P\right]^3\}  \nonumber \\
N_R^F&=&4 h_{(2,0)} + 4 \,\,\,\{- 4 h_{(1,0)}\}
\end{eqnarray}
where the terms in the curly brackets represent the effect of a
left--right symmetric gauging  of the $b_1$ non--dynamical gauge
fields. The coupling of the two sectors to these vector fields
reduces indeed the scalar degrees of freedom by $b_1$ and the
fermionic ones by $2\, b_1$.  This gauging is necessary in order to
restore supersymmetry on the right--moving sector, which otherwise
would not hold \cite{moa}, as it can be easily checked from
eqs. (\ref{bf}) using the property  $\int_{T_6}\,
\left[P\right]^3=6(h^{(2,0)}(P)-h^{(1,0)}(P)+1)$. From the above
results the central charge is easily computed on our solution to be
\begin{equation}
c_L=\mbox{Vol}(P)=6\, y^1y^2y^3= 6\, n^2 N_1 N_2 N_3
\label{cl}
\end{equation}
%%%%%%%%%%%%%%%%%%%%%%%%%%%%%%%%%%%%%%%%%%%%%%%%%%%%%%%%%%%%%%%%%%%%
\subsection{Computation of $h$.}
The excitation level $h$ is clearly the non--zero mode contribution to
the  total momentum  $L_0-\bar{L}_0$ along $S_1$,  denoted by $x_0$ in
table \ref{Mbrane2}.  We may therefore write $h=x_0-\Delta x_0$, where
$\Delta x_0$ is the  zero--mode contribution to the same momentum and
is the quantity which remains to be computed. 

It is instructive to
express $\Delta x_0$ in terms of type IIA quantities, going into a
regime ($R^6\ll \mbox{Vol}(T_6)$)  in which the low energy dynamics is no
more described by the $1+1$ SCFT but  by the zero modes of the open
strings attached to the three sets of $N_1,N_2,q^2\,N_3$ D4--branes
deriving from the dimensional reduction of the M5--branes and to the
$x_0=N_0+m^2\, N_3$ D0--branes on top of them, see table \ref{Mbrane2}. 
Before performing the
dimensional reduction, let us shift the eleven dimensional metric
$G_{MN}$ from the Minkowski background by an infinitesimal symmetric
matrix whose only non vanishing entries are those along the directions
$(0,10)$: $G_{MN}=\eta_{MN}+\delta G_{MN}$. In the low energy action
on the M5--brane world volume the  term $\delta S\propto \int_{M5}\,
T_{0\,10}\,\delta \,\hat{G}^{0\,10}$ would appear, representing the
contribution to the action of the  momentum along $S_1$. In the
following we shall be interested   in the contribution to the
energy--momentum tensor  $T_{nm}$ associated with the form $b^{(2)}$,
namely $T_{mn}(h)\propto\,  h^{(3)}_{mkl}\,{h^{(3)}_{n}}^{kl}$, which
encodes the zero mode contribution to be evaluated. After
compactifying on $S_1$ we obtain the type IIA configuration: the
deformation $\delta G_{10\,N}$ becomes the R--R one--form $C^{(1)}_N$
coupled to the D0--brane and the components $h^{(3)}_{10\,kl}$ give
rise to the vector field strength components on the D4 brane world
volumes, namely ${\cal F}_{kl}/(2\pi)$. The same action term $\delta
S$ on the D4--brane world volume can be shown to reduce  to: $\delta
S=-1/2(2\pi)^2 \left[\int_{P\times \IR}\, {\cal F}\wedge  {\cal
F}\wedge\, C^{(1)}\right]$. This is the Chern-Simons term defining the
effective D0--brane charge induced by a magnetic flux on the world
volumes of the D4--branes. \par In the regime of validity of the low
energy description of the system in terms of the $1+1$ SCFT the zero
mode contribution to $\delta S$ corresponds in the type IIA setting to
a magnetic flux ${\cal F}^{(0)}$ along the cycles $(4,5),\,
(6,7),\,(8,9)$  equal on the intersecting D4--branes along their
common directions:
\begin{eqnarray}
\Delta x_0=\frac{-1}{2(2\pi)^2}\int_P \,{\cal F}^{(0)}\wedge {\cal F}^{(0)}
\label{dx0}
\end{eqnarray}
The field strength ${\cal F}^{(0)}$ can be expanded along the cycles
$\alpha_i$ previously introduced: ${\cal F}^{(0)}={\cal
F}^{(0)|i}\,\alpha_i$.  The values of ${\cal F}^{(0)|i}$ are
determined in terms of the electric four dimensional charges $x_i$
from the Chern--Simons couplings of the magnetic flux to the R--R form
$C^{(3)}$ in the D4--brane world volume theory, and from the
requirement that the three matter vector potentials $A^i_\mu$ in the
STU model derive from the reduction of $C^{(3)}$ on $\alpha_i$:
$C^{(3)}=\sum_i\,A^i_\mu\, dx^\mu\wedge\alpha_i $.  The equations for
${\cal F}^{(0)|i}$ are: $x_i=6\, D_{ijk}\,y^j\,{\cal
F}^{(0)|k}/(2\pi)$. Solving them we are able to express  ${\cal
F}^{(0)|i}$, and therefore $\Delta x_0$, using eq. (\ref{dx0}), in
terms of $y^i$ and $x_i$: $\Delta x_0=m^2\,
N_3\,(N_1+N_2)^2/(4\,N_1\,N_2)$. The excitation level for our
configuration is therefore:
\begin{equation}
h=x_0-\Delta x_0=N_0+m^2\, N_3-m^2\, N_3\,\frac{(N_1+N_2)^2}{4\,N_1\,N_2}  
\label{n}
\end{equation}
implementing eqs. (\ref{cl}), (\ref{n}) in eq. (\ref{entropy}) we
obtain  precisely the expression for the macroscopic entropy in
eq. (\ref{Smicro})! This result provides a statistical interpretation
of Bekenstein--Hawking area law for the thermodynamical entropy whose validity,
for the very definition of generation solution, automatically extends
to the most general black hole of the same  kind.

%%%%%%%%%%%%%%%%%%%%%%%%%%%%%%%%%%%%%%%%%%%%%%%%%%%%%%%%%%%%%%%%%%%%%
\section{Discussion}
There are still many things to be understood on the front of BPS black
 holes. We have focused on a particular class of these solutions,
 namely the regular four dimensional BPS black holes in the $N=8$
 theory and addressed the question of how much may be learned already
 at the classical supergravity level on their microscopic
 description. This naturally  led to the study of the corresponding
 $U$--duality orbit and its generating solution. One of the main goals
 of our analysis in \cite{mm1,mm2,micro} is to have clarified the
 meaning of the five parameters characterizing this orbit (and
 therefore encoding all the microscopic degrees of freedom of a
 generic solution in it) both macroscopically and
 microscopically. The first macroscopic descriptions  of the
 generating solution were given in \cite{cvetic1,cvetic2,hull} within the
 framework of the heterotic string theory, nevertheless the
 microscopic interpretation of its parameters remained obscure (in
 particular of the parameter which lifted the solution
 from a purely dilatonic black hole to a generating solution). In
 \cite{bala} a suitable D--brane configuration was conjectured to
 describe the generating black hole,  but its macroscopic description
 as a R--R charged solution was missing. We have filled this gap by
 providing both the macroscopic and microscopic description of a
 particular generating solution \cite{mm2} and defining a framework in
 which to study all possible microscopic realizations of the solution
 and their duality relations at a classical supergravity level
 \cite{mm1}. In the light of these results  the mysterious fifth
 parameter, which in the description of table \ref{Mbrane2} is
 proportional to the parameter $m$, turns out to be related to a
 conserved quantity ($T_{10\,m}$ is indeed a conserved current on
 the M5 world-volume SCFT at least in a flat background) associated
 with the $b^{(2)}$ degrees of freedom in the M--theory picture or to
 a non trivial magnetic flux on the D4--brane world volumes in the
 type IIA picture. This charge allows the system to couple non--trivially 
to the axionic fields, which are naturally interpreted in
 the type IIA setting  as the components along the compact directions
 of the Kalb--Ramond field and in type IIB framework as off--diagonal
 components of the internal metric. Finally a further step along the microscopic analysis
 of the generating solution was achieved in \cite{micro} by computing its entropy from
 a microscopic state counting. This achievement, besides providing a
 statistical interpretation of the Bekenstein--Hawking entropy of  the
 most general regular BPS black hole in the theory under
 consideration, shed some light on how the five parameters enter the
 quantum  low energy dynamics on the generating solution and therefore
 the  expression of the microscopic entropy itself.

\vskip 1cm
\noindent 
{\bf Acknowledgements}

Different sections of this paper have to appear in the
proceedings for: the IX Marcel Grossman Meeting (Rome, July 2000) , the
TMR ``Integrability Nonperturbative Effects And Symmetry In Quantum
Field Theory'' conference (Paris, September 2000), and the RTN ``The
Quantum Structure of  Spacetime and the Geometric Nature of
Fundamental Interactions''' conference (Berlin, October 2000). For
sake of economy and completeness and not to bother the potential reader, we have 
unified and homogenized the different parts in the version for the archive. 
This work is partially supported by the European Commission RTN programme 
HPRN--CT--2000--00131. M.B. acknowledges support by INFN.

\end{document}